\documentclass[aps,pra,twocolumn,floatfix]{revtex4}

\usepackage{graphicx}
\usepackage{color}
\begin{document}
\title{Molecular states near a collision threshold}
\author{Paul S. Julienne}
\affiliation{Joint Quantum Institute, National Institute of
Standards and Technology and University of Maryland, 100 Bureau Drive Stop 8423, Gaithersburg, Maryland 20899-8423, USA}

\date{\today}

\begin{abstract}
[To appear as Chapter 6 of { \it Cold Molecules: Theory, Experiment, Applications}, ed. by Roman Krems et al. (Taylor and Francis, 2009)]
\end{abstract}

\maketitle

\section{Introduction} \label{sec:psj_intro}

Real atoms are typically complex, having ground and excited states with spin structure.  The molecules formed from the atoms typically have a rich spectrum of near-threshold bound and quasi-bound molecular states when the molecular spin, rotational, and vibrational structure is taken into account.   When an ultracold gas of atoms is produced, the atoms are prepared in specific quantum states, and collisions between the atoms occur with an extremely precisely defined energy close to the $E=0$ collision threshold of the interacting atoms, where $E$ denotes energy.  The collision then makes the near-threshold spectrum of the molecular complex of the two atoms accessible to electromagnetic probing.   An external magnetic or electromagnetic field can be precisely tuned to couple the colliding atoms to a specific molecular state, which can be viewed as a scattering resonance.  This permits both extraordinary spectral accuracy in probing near-threshold level positions (order of $E/h=10$ kHz accuracy for 1 $\mu$K atoms) and precise resonant control of the collisions that determine both static and dynamical macroscopic properties of quantum gases.  Consequently, understanding the near-threshold bound and scattering states is essential for understanding the collisions and interactions of ultracold atoms.  This is also true for interactions of ultracold molecules.

This Chapter concentrates on understanding molecules that can be made by combining two cold atoms using either magnetically tunable Feshbach resonance states~\cite{Kohler06} or optically tunable photoassociation resonance states~\cite{Jones06}. Such resonances provide a mechanism for the formation of ultracold molecules from already cold atoms.  In addition, magnetically tunable resonances have been used very successfully to control the properties of ultracold quantum gases.  This Chapter treats both magnetically and optically tunable molecular resonances with the same scattering theory framework.  The viewpoint from quantum defect theory is emphasized of conceptually separating the interaction of the atoms into short range and long range regions.   These regions are characterized by very different energy and length scales.    Much insight about near-threshold collisions and bound states, as well as practical tools for their study, can be gained by taking advantage of this separation~\cite{Julienne89,Moerdijk95,Vogels98,Burke98,Vogels00,Mies00,Gao00,Gao01,Julienne06}.  While molecular physics is typically concerned with strong short range interactions associated with "ordinary molecules," ultracold physics is concerned with scattering states and very weakly bound molecular states in the threshold domain near $E=0$.  The long range potential, which has a lead term that varies as $1/R^n$, plays an important role in connecting  these two regimes.

We briefly summarize here the theory of cold collisions, which is described in detail in Chapter {\color{red} XXX}.  The scattering wavefunction is expanded in states of relative angular momentum of the two atoms characterized by partial wave quantum number $\ell = 0,1,2\ldots$.  Generally, the atoms can be initially prepared in one of several quantum states, and the scattering "channels" can be specified by a collective set of quantum numbers $\alpha$ representing the state of each atom and the partial wave.  Upon solving the Schr{\"o}dinger equation for the system, the effect of all short-range interactions during a collision with $E>0$ is summarized in the scattering wavefunction  for $R \to \infty$ by a unitary $S$-matrix.  Only the lowest few partial waves can contribute to cold collisions, and in the limit $E \to 0$, only $s$-wave channels with $\ell=0$ have non-negligible collision cross sections.  Using the complex scattering length $a-ib$ to represent the $s$-wave $S$-matrix element $S_{\alpha\alpha}=\exp{[-2ik(a-ib)]}$ in the limit $E \to 0$, the contribution to the elastic scattering cross section from $s$-wave collisions in channel $\alpha$ is
\begin{equation}
\label{eq:psj_sigmael}
 \sigma_\mathrm{el} = \lim_{E\to 0} g\frac{\pi}{k^2}\left | 1 - S_{\alpha\alpha} \right |^2
  = 4g\pi\left ( a^2+b^2\right ) \,,
\end{equation}
where $\hbar k = \sqrt{2\mu E}$ is the relative collision momentum in the center of mass frame for the atom pair with reduced mass $\mu$.  The rate coefficient $K_\mathrm{loss}=\sigma_\mathrm{loss} v$ for $E \to 0$ $s$-wave inelastic collisions that remove atoms from channel $\alpha$ is 
\begin{equation}
\label{eq:psj_Kloss}
K_\mathrm{loss} = \lim_{E\to 0} g\frac{\pi\hbar}{\mu k} \left ( 1 - |S_{\alpha\alpha}|^2 \right ) =
2g\frac{h}{\mu}  b
\end{equation}
where $v=\hbar k/\mu$ is the relative collision velocity.  The symmetry factor $g=1$ when the atoms are bosons or fermions that are not in identical states, $g=2$ or $g=1$ respectively for two bosons in identical states in a normal thermal gas or a Bose-Einstein condensate, and $g=0$ for two fermions in identical states.  If there are no exoergic inelastic channels present, then $b=0$ and only elastic collisions are possible.

The Schro{\"o}dinger equation also determines the bound states with discrete energies $E_i<0$.  While the conventional picture of molecules counts the bound states by vibrational quantum number $v=0,1\ldots$ from the lowest energy ground state up, it is more helpful for the present discussion to count the near-threshold levels from the $E=0$ dissociation limit down by quantum numbers $i=-1,-2\ldots$.  In the special case where $a \to +\infty$, the energy of the last bound $s$-wave state of the system with $i=-1$ depends only on $a$ and $\mu$ and takes on the following "universal" form:
\begin{equation}
\label{eq:psj_universalE}
 E_{-1} = -\frac{\hbar^2}{2\mu a^2} \,\,\mathrm{as}\,\,a\to +\infty\,.
\end{equation}

Section~\ref{sec:psj_singleV} describes the bound and scattering properties of a single potential with a van der Waals long range form.  Section~\ref{sec:psj_multiV} extends the treatment to multiple states and scattering resonances.  Sections~\ref{sec:psj_magres} and~\ref{sec:psj_optres} respectively discuss the properties of magnetically and optically tunable molecular resonance states.

\section{Properties for a single potential} \label{sec:psj_singleV}

\begin{figure}[tbp]
\centering
\includegraphics[width=\columnwidth]{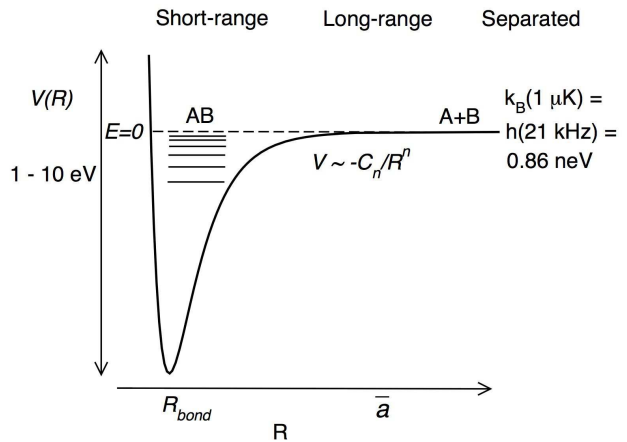}
\caption{Schematic figure of the potential energy curve $V(R)$ as a function of the separation $R$ between two atoms $A$ and $B$.  The horizontal lines labeled $AB$ indicate a spectrum of molecular bound states leading up to the molecular dissociation limit at $E=0$, indicated by the dashed line.  The long range potential varies as $-C_n/R^n$. } \label{fig:psj1}
\end{figure}

In this section let us ignore any complex internal atomic structure and first consider two atoms $A$ and $B$ that interact by a single adiabatic Born-Oppenheimer interaction potential $V(R)$, illustrated schematically in Fig.~\ref{fig:psj1}.   The wavefunction for the system is $|\alpha\rangle |\psi_\ell\rangle/R$, where $|\alpha\rangle$ represents the electronic and rotational degrees of freedom, and the wavefunction for relative motion is found from the radial Schr{\"o}dinger equation
\begin{equation}
 -\frac{\hbar^2}{2\mu}\frac{d^2 \psi_\ell}{dR^2} +\left (V(R)+\frac{\hbar^2\ell(\ell+1)}{2\mu R^2} \right ) \psi_\ell = E \psi_\ell \,. 
 \label{eq:psj_SE}
\end{equation}
Solving Eq.~\ref{eq:psj_SE} gives the spectrum of bound molecular states $\psi_{i\ell}$ with energy $E_{i\ell}=-\hbar^2 k_{i\ell}^2/(2\mu)<0$ and the scattering states $\psi_\ell(E)$ with collision kinetic energy $E=\hbar^2 k^2/(2\mu)>0$, where $k_{i\ell}$ and $k$ have units of $(\mathrm{length})^{-1}$.  As $R \to \infty$, the bound states decay as $e^{-k_{i\ell} R}$ and the scattering states approach
\begin{equation}
 \psi_\ell(E) \to c \sin(kR-\pi \ell/2+\eta_\ell)/k^{1/2} \,. 
 \label{eq:psj_psi}
\end{equation}
 Bound states are normalized to unity, $|\langle \psi_{i\ell}|\psi_{j\ell'}\rangle|^2 =\delta_{ij}\delta_{\ell\ell'}$.  We choose the normalization constant $c=\sqrt{2\mu / \hbar^2 \pi}$ so that scattering states are normalized per unit energy, $\langle \psi_\ell(E)|\psi_{\ell'}(E')\rangle=\delta(E-E')\delta_{\ell\ell'}$.  Thus, the energy density of states is included in the wavefunction when taking matrix elements involving scattering states.

The long range potential between the two atoms varies as $-C_n/R^n$.  We are especially interested in the case of $n=6$ for the van der Waals interaction between two neutral atoms.  This is the lead term in the long-range expansion of the potential in inverse powers of $R$ that applies to many atoms that are used in ultracold experiments.  This potential has a characteristic length scale of $R_\mathrm{vdw}=\sqrt[4]{2 \mu C_6/\hbar^2}/2$ that depends only on the values of $\mu$ and $C_6$~\cite{Jones06}.  Values of $C_6$ are tabulated by Derevianko~\cite{Derevianko99} for alkali-metal species and by Porsev and Derevianko~\cite{Porsev06} for alkaline-earth species.  We prefer to use a closely related van der Waals length introduced by Gribakin and Flambaum~\cite{Gribakin93}
\begin{equation}
\bar{a}=4\pi/\Gamma(1/4)^2\,R_\mathrm{vdw}=0.955978\dots\,R_\mathrm{vdw}\,,
\label{eq:psj_abar}
\end{equation}
where $\Gamma(x)$ is the Gamma function.   This length defines a corresponding energy scale $\bar{E}=\hbar^2/(2\mu \bar{a}^2)$.  The parameters $\bar{a}$ and $\bar{E}$ occur frequently in formulas based on the van der Waals potential.   The wavefunction approaches its asymptotic form when $R\gg \bar{a}$ and is strongly influenced by the potential when $R \lesssim \bar{a}$.  Table~\ref{tab:psj_vdw} gives the values of $\bar{a}$ and $\bar{E}$ for several species used in ultracold experiments.

\begin{table}
\caption{ Characteristic van der Waals scales $\bar{a}$ and $
\bar{E}$ for several atomic species. (1 amu = 1/12 mass of a
$^{12}$C atom, 1 au= 1 $E_h a_0^6$ where $E_h$ is a hartree and 1
$a_0$= 0.0529177 nm) } \label{tab:psj_vdw}
\begin{tabular}{ccccccl}
\hline\hline
Species & mass & C$_6$  & $\bar{a}$ & $\bar{E}/h$ & $\bar{E}/k_B$\\
     & (amu) & (au) & ($a_0$) & (MHz) &  (mK) \\
\hline
${^6}$Li &      6.015122 & 1393 & 29.88 & 671.9 &  32.25\\
${^{23}}$Na & 22.989768 & 1556 & 42.95 & 85.10 & 4.084\\
${^{40}}$K & 39.963999 & 3897 & 62.04 & 23.46 &  1.126\\
${^{87}}$Rb & 86.909187 & 4691 & 78.92 &  6.668 & 0.3200\\
$^{88}$Sr &  87.905616 & 3170 & 71.76 & 7.974 & 0.3827\\
${^{133}}$Cs & 132.905429 &  6860 & 96.51 & 2.916 & 0.1399\\
${^{174}}$Yb & 173.938862  & 1932 &  75.20 &  3.670 & 0.1761\\
\hline\hline
\end{tabular}\\
\end{table}

Samples of cold atoms can be prepared with kinetic temperatures on the order of nK to mK.  The energy associated with temperature $T$ is $k_B T$ where $k_B$ is the Boltzmann constant.  For example, at $T=1$ $\mu$K, $k_B T =0.86$ neV and $k_B T/h=21$ kHz.  This ultracold energy scale is 9 to 10 orders of magnitude smaller than the energy scale of 1 to 10 eV associated with ground or excited state interaction energies when a molecule is formed at small interatomic separation $R_\mathrm{bond}$ on the order of a chemical bond length.   In a cold collision, the initially separated atoms have very low collision energy $E=\hbar^2 k^2/(2\mu)\approx 0$ and very long de Broglie wavelength $2\pi/k$.  The atoms come together from large distance $R$ and are accelerated by the interatomic potential $V(R)$, so that when they reach distances on the order of $R_\mathrm{bond}$  they have very high kinetic energy on the order of $|V(R_\mathrm{bond})|$.  The local de Broglie wavelength $2\pi/k(R,E)$ in the short range classical part of the potential, where $k(R,E)=\sqrt{2\mu (E-V(R))}/\hbar$, is orders of magnitude smaller than the separated atom de Broglie wavelength and is nearly independent of the value of $E$, which is close to $0$.

\begin{figure}[tbp]
\centering
\includegraphics[width=\columnwidth]{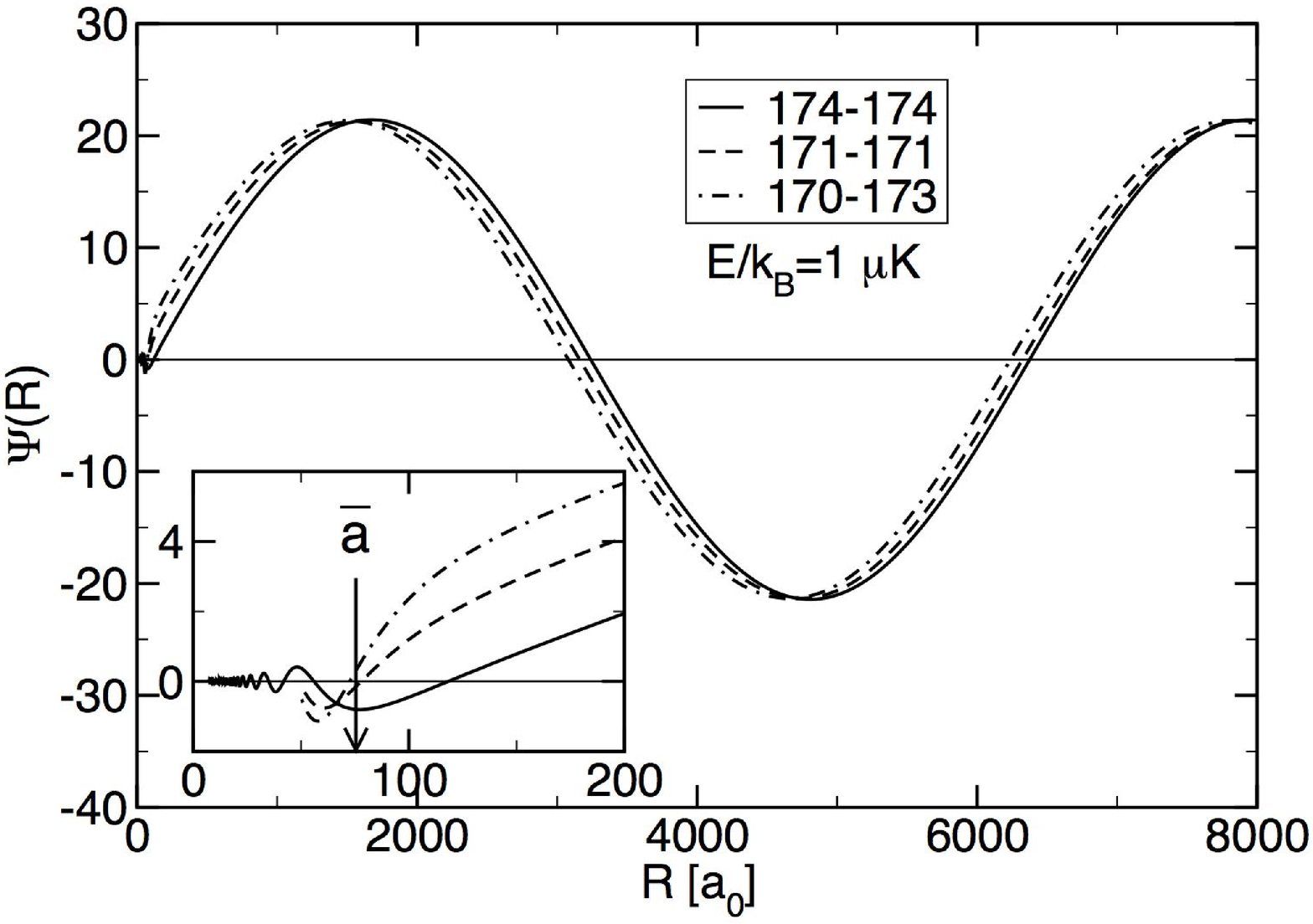}
\caption{Radial wavefunction $\psi_0(R)$ for $\ell=0$ at $E/k_b=1$ $\mu$K for the pairs $^{174}$Yb-$^{174}$Yb (solid), $^{171}$Yb-$^{171}$Yb (dashed), and $^{170}$Yb-$^{173}$Yb (dotted), which have respective scattering lengths of 105 a$_0$, -3 a$_0$, and $-81$ a$_0$~\cite{Kitagawa07}.  The inset shows an expanded view of the wavefunction on a smaller length scale on the order of $\bar{a}$, the characteristic length of the van der Waals potential.  The $^{174}$Yb-$^{174}$Yb  case  shows the oscillations that develop when $R < \bar{a}$.} \label{fig:psj2}
\end{figure}

This separation of scales is illustrated in Fig.~\ref{fig:psj2}, which shows examples of $s$-wavefunctions at a collision energy $E/k_B=$ 1 $\mu$K, where dividing $E$ by $k_B$ allows us to express energy in temperature units.  This example uses three isotopic combinations of pairs of Yb atoms, which has a spinless $^1$S$_0$ electronic configuration and a single ground state electronic Born-Oppenheimer potential $V(R)$.  The species Yb makes a good example case to illustrate the principles in this section, since it has 7 stable isotopes and 28 different atom pairs of different isotopic composition for which the threshold properties have been worked out~\cite{Kitagawa07}.    All combinations have the same $V(R)$ but different reduced masses.  This mass-scaling approximation, which ignores very small mass-dependent corrections to the potential, is normally quite good except for very light species such as Li.  Fig.~\ref{fig:psj2} shows that the three examples have similar phase-shifted sine waves with a common long de Broglie wavelength of $2\pi/k=6300$ a$_0$.  For small $R$ where $kR\ll 1$ the sine function vanishes as $c \sin k(R-a)/\sqrt{k} \to c \sqrt{k}(R-a)$.  The actual wavefunction oscillates rapidly at small $R$ due to the influence of the potential.  Since the asymptotic form for $kR \ll 1$ varies as $k^{1/2}$ as $k\to 0$, the short range oscillating part also has an amplitude proportional to $k^{1/2}$ in order to connect smoothly to the asymptotic form as $k\to 0$.  This property ensures that the threshold matrix elements that characterize Feshbach resonances and $s$-wave  inelastic scattering are proportional to $k^{1/2}$.

\begin{figure}[tbp]
\centering
\includegraphics[width=\columnwidth]{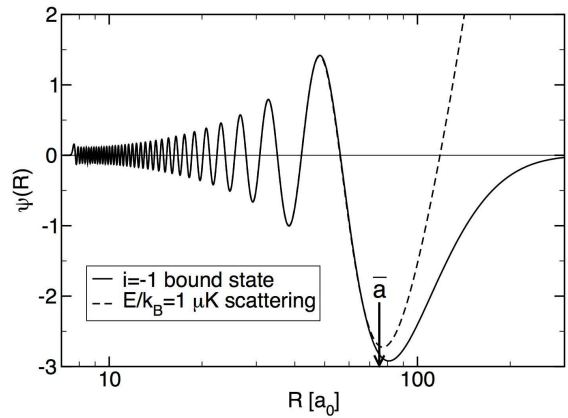}
\caption{wavefunctions for the last $s$-wave $i=-1$ bound state (solid line) with $E_{-1,0}/h=-10.6$ MHz and for the $s$-wave scattering state (dashed line) for $E/h=0.02$ MHz ($E/k_B=1$ $\mu$K) for two $^{174}$Yb atoms.  Both wavefunctions are given a common JWKB normalization at small $R \ll \bar{a}$ and are nearly indistinguishable for $R < \bar{a}$.  The potential supports $N=72$ bound states, and the wavefunction for this $i=-1$ and $v=71$ level has $N-1=71$ nodes. } \label{fig:psj3}
\end{figure}

Figure~\ref{fig:psj3} illustrates more clearly the nature of threshold short range scattering and bound state wavefunctions.  When given an appropriate short range normalization, near-threshold scattering and bound state wavefunctions have a common amplitude and phase in the region of $R$ small compared to the range $\bar{a}$ of the long range potential.  While this can be put on a rigorous quantitative ground within the framework of quantum defect theory~\cite{Mies00}, it is easy to show using the familiar JWKB approximation~\cite{Julienne89,Vogels00,Jones06}.   We can always write the wavefunction in phase-amplitude form $\psi_{\ell}(R,E) = \alpha_\ell(R,E)\sin{\beta_\ell(R,E)}$ and transform the Schr{\"o}dinger equation~(\ref{eq:psj_SE}) into a set of equations for $\alpha_\ell$ and $\beta_\ell$.  The asymptotic $\psi_\ell(R,E)$ in Eq.~(\ref{eq:psj_psi}) clearly corresponds to this form with $\alpha_\ell \to c/k^{1/2}$ as $R \to \infty$.  Another familiar form is the JWKB semiclassical wavefunction $\psi_\ell^{JWKB}(R,E)$, for which
\begin{eqnarray}
\alpha_\ell^{JWKB}(R,E)   & = & c/k_\ell(R,E)^{1/2}   \label{eq:psj_alpha}\\
\beta_\ell^{JWKB}(R,E) & = & \int_{R_t}^R k_\ell(R',E)dR'  + \frac{\pi}{4} \label{eq:psj_beta} \,.
\end{eqnarray}
where $R_t$ is the inner classical turning point of the potential.

When the collision energy $E$ is sufficiently large, so there are no threshold effects, the JWKB approximation is a excellent approximation at all $R$, and the form of $\alpha_\ell^{JWKB}(R,E)$ in Eq.~(\ref{eq:psj_alpha}) applies at all $R$, transforming into the correct quantum limit as $R\to \infty$.  On the other hand, the JWKB approximation fails for $s$-waves with very low collision energy. This failure occurs in a region of $R$ near $\bar{a}$ and for collision energies $E$ on the order of $\bar{E}$ or less.  The consequence is that the JWKB wavefunction, with the normalization in Eq.~(\ref{eq:psj_alpha}), is related to the actual wavefunction, with the asymptotic form in Eq.~(\ref{eq:psj_psi}), by a multiplicative factor $C_\ell(E)$, so that as $E \to 0$
\begin{equation}
\label{eq:psj_psiE+}
 \psi_\ell(R,E) = C_\ell(E)^{-1} \psi_\ell^{JWKB}(R,0) \,.
\end{equation}
As $k\to 0$ for a van der Waals potential varying as $1/R^6$, the $s$-wave threshold form is $C_0(E)^{-2}=k\bar{a} [1 + (r-1)^2]$, where $r=a/\bar{a}$ is the dimensionless scattering length in units of $\bar{a}$~\cite{Mies00}.  Equation~(\ref{eq:psj_psiE+}) gives an excellent approximation for the threshold $\psi_0(R,E)$ for $R < \bar{a}$ and $k < 1/a$.    At high energy, when $E \gg \bar{E}$, $C_0(E)^{-1}$ approaches unity and the JWKB approximation for $\psi_0(R,E)$ applies at all $R$.

The unit normalized bound state wavefunction $\psi_{i\ell}(R)$ can be converted to an "energy normalized" form by multiplying by $|\partial i/\partial E_{i\ell}|^{1/2}$, where $-\partial i/\partial E_{i\ell}>0$ is the energy density of states.  Away from threshold, this is just the inverse of the mean spacing between levels, whereas for $s$-wave levels near threshold for a van der Waals potential , $\partial i/\partial E_{i0} \to r/(2\pi\bar{E})^{-1}$ as $k_{-1,0}=1/a  \to 0$~\cite{Mies00}.  The relation of  $\psi_{i\ell}$ to the energy-normalized JWKB form in the classically allowed region of the potential is
\begin{equation}
\label{eq:psj_psi-}
 \psi_{i\ell}(R,E_{i\ell}) = \left | \frac{\partial i}{\partial E_{i\ell}} \right |_{E_{i\ell}}^{-1/2} \psi_\ell^{JWKB}(R,E_{i\ell}) \,,
\end{equation}

Figure~\ref{fig:psj3} plots $C_0(E)\psi_0(R,E)\approx \psi_0^{JWKB}(R,0)$ for the scattering state and  $|\partial i/\partial E_{i0}|^{1/2} \psi_{i0}(R,E_{i0}) \approx \psi_0^{JWKB}(R,0)$ for the $i=-1$ bound state.  Thus the near-threshold bound and scattering wavefunctions, when given a common short range normalization, are nearly identical and are well approximated by $\psi_0^{JWKB}(R,0)$ in the region $R < \bar{a}$.  For $R>\bar{a}$ the wavefunctions begin to take on their asymptotic form as $R\to \infty$.  The shape of the wavefunction at very small $R$ on the order of $R_\mathrm{bond}$ is usually independent of $E$ for ranges of $E/k_B$ on the order of many K.  The short range shape  is even independent of $\ell$ for small $\ell$, since the rotational energy is very small compared to typical values of $V(R_\mathrm{bond})$.  However, the {\em amplitudes} of the wavefunctions depend strongly on the whole potential, which determines $a$, and are analytically related to the form of the long range potential.  

The separation of scales for $R>\bar{a}$ and $R<\bar{a}$ is a key feature of ultracold physics that enables much physical insight as well as practical approximations to be developed about molecular bound and quasibound states and collisions.    Given that $C_6$, $\mu$, and the $s$-wave scattering length $a$ are known, the Schr{\"o}dinger equation~(\ref{eq:psj_SE}) can be integrated inward using the form of Eq.~(\ref{eq:psj_psi}) as $k \to 0$ as a boundary condition, thus giving the wavefunction and nodal pattern for $R<\bar{a}$ as $E \to 0$.   Assume that it is possible to  pick some $R=R_m$ such that $R_\mathrm{bond}\ll R_m \ll \bar{a}$ and  $V(R_m)$ is well-represented by its van der Waals form.  Then the log of the derivative of the wavefunction at $R_m$, which also can be calculated, provides an inner boundary condition, independent of $E$ over a wide range of $E$, for matching the wavefunction at $E$ propagated from large $R$.  All that is needed to do this is to know $a$, $\mu$ and the long range potential.  Thus, it is readily seen that all of the near threshold bound and scattering states, even those for $\ell>0$, can be calculated to a very good approximation for $R>R_m$ once $C_6$, $\mu$, and $a$ are known.

\begin{figure}[tbp]
\centering
\includegraphics[width=\columnwidth]{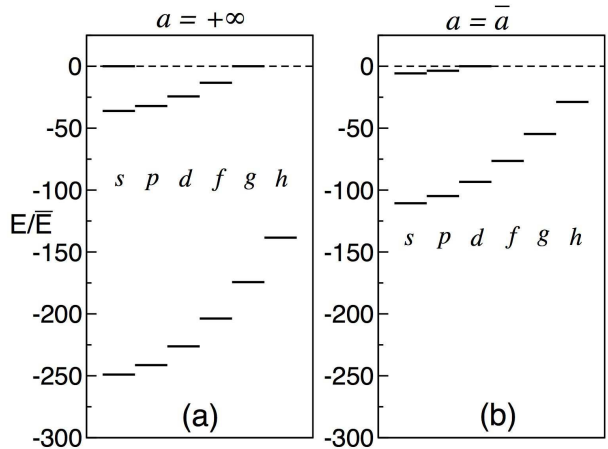}
\caption{Dimensionless bound state energies $E_{i\ell}/\bar{E}$ for partial waves $\ell=0 \ldots 5$ $(s,p,d,f,g,h)$.  Panel (a) is for the case $a=\pm \infty$ and Panel (b)  is for $a=\bar{a}$.}
\label{fig:psj4}
\end{figure}

Figure~\ref{fig:psj4} shows the spectrum of bound states $E_{i\ell}$, in units of $\bar{E}$, for $\ell$ up to 5 for two cases of scattering length, based on the van der Waals quantum defect theory of Gao~\cite{Gao00,Gao01}.  Panel (a) shows the case of $a=\pm\infty$, where there is a bound state at $E=0$.  The locations of the bound states for $a=\pm\infty$ define the boundaries of the "bins" in which, for any $a$, there will be one and only one $s$-wave bound state, for example, $-36.1\bar{E} < E_{-1,0}<0$ and $-249\bar{E}<E_{-2,0}<-36.1\bar{E}$.  The panel also shows the rotational progressions for each level as $\ell$ increases.  The $a=\pm\infty$ van der Waals case also follows a "rule of 4", where partial waves $\ell=4,8,\ldots$ also have a bound state at $E=0$.  Panel (b) shows how the spectrum changes when $a=\bar{a}$, for which the there is a $d$-wave level at $E=0$.  Similar spectra can be calculated for any $a$.  

Gribakin and Flambaum~\cite{Gribakin93} showed that the near-threshold $s$-wave bound state for a van der Waals potential in the limit $a \gg \bar{a}$ is modified from the universal form in Eq.~(\ref{eq:psj_universalE}) as 
\begin{equation}
\label{eq:psj_vdwE}
 E_{-1} = -\frac{\hbar^2}{2\mu (a-\bar{a})^2}\,.
\end{equation}
This approaches the universality limit when $a \gg \bar{a}$, in which case the $s$-wave wavefunction takes on the universal form $\psi_0(R,E) = \sqrt{2/a} e^{-R/a}$.  Such an exotic bound state, known as a ``halo molecule,'' exists primarily in the nonclassical domain beyond the outer classical turning point of the long-range potential with an expectation value of $R$ of $a/2$, which grows without bound as $a \to +\infty$~\cite{Kohler06}.

\begin{figure}[tbp]
\centering
\includegraphics[width=\columnwidth]{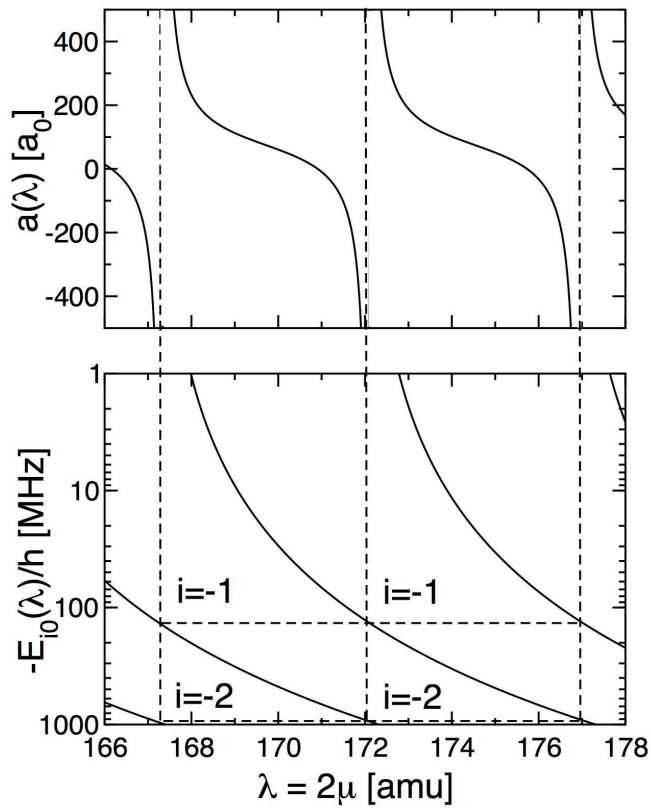}
\caption{The upper panel shows $s$-wave scattering length and the lower panel shows bound state binding energies $-E_{i0}(\lambda)$ for Yb$_2$ molecular dimers versus the control parameter $\lambda=2\mu$.  The vertical dashed lines show the points of singularity of $a(\lambda)$.  The horizontal dashed lines show the boundaries of the bins in which the $i=-1$ and $i=-2$ levels must lie.} \label{fig:psj5}
\end{figure}

Bound state and scattering properties are closely related.  It is instructive to imagine that there is some control parameter $\lambda$ that can be varied  to make the scattering length vary over its whole range from $+\infty$ and $-\infty$, changing the corresponding bound state spectrum.  One way to do this would be to vary the reduced mass.  Of course, this is not physically possible.  However, there are elements with many isotopes, so that a wide range of discrete reduced masses are possible.  An excellent physical system to illustrate this is the Ytterbium atom, used in the examples of Figs.~\ref{fig:psj2} and \ref{fig:psj3}.  The stable isotopes with masses 168, 170, 172, 174, and 176 are all spinless bosons and the 171 and 173 isotopes are fermions with spin $1/2$ and $5/2$ respectively.    Yb atoms can be cooled into the $\mu$K domain and all isotopes, including the fermionic ones in different spin states, have s-wave interactions..  The locations of several $\ell=0$ and 2 threshold bound states of different isotopic combinations of Yb atoms in Yb$_2$ dimer molecules have been measured, and the long range potential parameters and scattering lengths determined~\cite{Kitagawa07}.

Figure~\ref{fig:psj5} shows the $s$-wave scattering length and  bound state binding energies versus the continuous control parameter $\lambda =2\mu$.  Physically, there are 28 discrete values between $\lambda=$168 and 176.  The scattering length has a singularity, and a new bound state occurs with increasing $\lambda$, at $\lambda=$ 167.3, 172.0, and 177.0.  The range between 167.3 and 172 corresponds to exactly $N=71$ bound states in the model potential used.   Near $\lambda=167.3$ the last $s$-wave bound state energy $E_{-1,0} \to 0$ as $-\hbar^2/(2\mu a^2)$ as $a \to +\infty$.  The binding energy $|E_{-1,0}|$ gets larger as $\lambda$ increases and $a$ decreases, so that for a van der Waals potential $E_{-1,0}$ approaches the lower edge of its "bin" at $-36.1 \bar{E}$ as $a \to -\infty$.  As $\lambda$ increases beyond $172.0$, the $i=-1$ level becomes the $i=-2$ level as a new $i=-1$``last'' bound state appears in the spectrum.

The variation of scattering length with $2\mu$ is given by a remarkably simple formula.  While semiclassical theory breaks down at threshold, Gribakin and Flambaum~\cite{Gribakin93} showed that the correct quantum mechanical relation between $a$ and the potential is
\begin{equation}
\label{eq:psj_aGF}
a = \bar{a} \left [ 1 - \tan{\left (\Phi-\frac{\pi}{8}\right )} \right ]\,,
\end{equation}
 where
 \begin{equation}
 \Phi=\int_{R_t}^\infty \sqrt{-2\mu V(R)/\hbar^2} = \beta_0^{JWKB}(\infty,0)-\pi/4\,.
 \label{eq:psj_Phi}
 \end{equation}
The number of bound states in the potential is $N=[\Phi/\pi - 5/8]+1$, where $[\ldots ]$ means the integer part of the expression.  These expressions work remarkably well in practice.  Although the results in Fig.~\ref{fig:psj5} are obtained by solving the Schr{\"o}dinger equation for a realistic potential, virtually identical results are obtained for $a$ from Eq.~(\ref{eq:psj_aGF}).  In fact, $a$ and $E_{i0}$ are nearly the same on the scale of Fig.~\ref{fig:psj5} if the simple hard-core van der Waals model of~\cite{Gribakin93} is used for the potential, namely $V(R)=-C_6/R^6$ if $R\ge R_0$ and $V(R)=+\infty$ if $R<R_0$, where the cutoff $R_0$ is chosen to fit $a$ or $E_{-1,0}$ data from two different isotopes.  With the mass scaling $\propto \sqrt{\mu}$ in Eq.~(\ref{eq:psj_Phi}), knowing $C_6$ and $E_{-1,0}$ for two isotopic pairs determines  $a$ and $E_{-1,0}$ for all isotopic pairs.  The approximation is fairly good even for levels with larger $|i|$ or $\ell>0$, although it will become worse as $|i|$ or $\ell$ increase. 

In summary, it is very useful to take advantage of the enormous difference in energy and length scales associated with the cold separated atoms and deeply bound molecular potentials.  This allows us to introduce a generalized ``quantum defect" approach for understanding threshold physics~\cite{Julienne89,Burke98,Gao01,Vogels00,Mies00}.  Threshold bound state and scattering properties are determined mainly by the  long range potential, once the overall effect of the whole potential is known through the $s$-wave scattering length.  A similar analysis can be developed for other long range potential forms, for example, $1/R^4$ ion-induced dipole or $1/R^3$ dipole-dipole interactions.

\section{Interactions for multiple potentials} \label{sec:psj_multiV}

Generally the cold atoms used in experiments have additional angular momenta (electron orbital and/or electron spin and/or nuclear spin), so that more than one scattering channel $\alpha$ can be involved in a collision.  Each channel has a separated atom channel energy $E_\alpha$.   Fig.~\ref{fig:psj1} could be modified to illustrate such channels by adding additional potentials and their corresponding spectra dissociating to the $E_\alpha$ limits.  If  $E_{tot}$ is the total energy of the colliding system, the designation open or closed is used for channels with $E_{tot} >E_\alpha$ or  $E_{tot}<E_\alpha$ respectively.  Inelastic collisions from entrance channel $\alpha$ are possible to open exit channels $\beta$ when $E_\alpha>E_\beta$, whereas closed channels $\beta$ can support quasibound states as scattering resonances when $E_\alpha<E_{tot}<E_\beta$.  The ability to tune resonance states to control scattering properties or to convert them into true molecular bound states is an important aspect of ultracold physics that has been exploited in a wide variety of experiments with bosonic or fermionic atoms~\cite{Kohler06}.

Let us first examine the basic magnitude of the $s$-wave inelastic collision rates that are possible when open channels are present.  The rate constant is  determined by the magnitude of $b$ in Eq.~(\ref{eq:psj_Kloss}), for which a typical order of magnitude is $b\approx \bar{a}$ for an allowed transition, that is, one with a relatively large short-range interactions in the system Hamiltonian.  The rate constant can be written
\begin{equation}
K_\mathrm{loss}= 0.84\times 10^{-10} g \frac{b[\mathrm{au}]}{\mu[\mathrm{amu}]}\,\,\mathrm{cm}^3/\mathrm{s}\,,
\label{eq:psj_Knum}
\end{equation}
where $b$ is expressed in atomic units (1 au $=$ 0.0529177 nm) and $\mu$ in atomic mass units ($\mu=12$ for $^{12}$C).  Allowed processes will typically have the order of magnitude of 10$^{-10}$ cm$^3/$s for $K_\mathrm{loss}$.  The $s$-wave $K_\mathrm{loss}$ can be even larger, with an upper bound of $b_u=1/(4k)$ being imposed by the unitarity property of the $S$-matrix, i. e., $0 \le 1-|S_{\alpha \alpha}|^2 \le1$.  Since the lifetime relative to  collision loss is $\tau = 1/(K_\mathrm{loss}n)$, where $n$ is the density of the collision partner, allowed processes result in fast loss with $\tau \lesssim1$ ms at typical quantum degenerate gas densities.  This applies to atom-molecule and molecule-molecule collisions as well as atom-atom collisions.  Such losses need to be avoided by working with atomic or molecular states that do not experience fast loss collisions, such as the lowest energy ground state level, which does not have exoergic 2-body exit channels.  Alternatively, placing the species in a lattice cell that confines a single atom or molecule can offer protection against collisional loss.

An alternative formulation of the collision loss rate is possible by rewriting Eq.~(\ref{eq:psj_Kloss}), not taking the $E \to 0$ limit but introducing a thermal average over a Maxwellian distribution of collision energies $E$,
\begin{equation}
\label{eq:psj_Ktherm}
 K_{loss} =g \frac{1}{Q_T} \frac{k_B T}{h} \sum_\alpha \left  \langle 1 - |S_{\alpha\alpha}|^2 \right \rangle_T \,,
\end{equation}
where $Q_T$ is the translational partition function, $1/Q_T=(2\pi\mu k_B T/h^2)^{3/2}=\Lambda_T^3$ where $\Lambda_T$ is the molecular thermal de Broglie wavelength.  The $\langle \ldots\rangle_T$ expression implies a thermal average over the velocity distribution.  The sum represents a dynamical factor $f_D$ that varies as $T^{1/2}$ as $T \to 0$ and has an upper bound of unity for $s$-waves and $\approx \ell_\mathrm{max}^2$ if $\ell_\mathrm{max}$ partial waves contribute at the unitarity limit.  Although Eq.~(\ref{eq:psj_Ktherm}) reduces to Eq.~(\ref{eq:psj_Knum}) in the $T\to 0$ $s$-wave limit, it lets us see that the collision rate is given by an expression having the form
\begin{equation}
\label{eq:psj_tau}
 \tau^{-1} = K_{loss}n = g (n\Lambda_T^3) \frac {k_B T}{h} f_D \,.
\end{equation}
This form embodies some general principles for any collisions of atoms and molecules.  The dimensionless $n\Lambda_T^3$ factor shows that the collision rate is proportional to phase space density of the collision partner (scale by mass ratios to convert to an atomic phase space density).  The $k_B T/h$ factor sets an intrinsic rate scale (dimension of inverse time) associated with $T$.  The dimensionless factor $f_D$ embodies all of the detailed collision dynamics.  Even using fast time-dependent manipulations to control $f_D$ does not change the fundamental thermodynamic limits imposed by the phase space density and $k_B T/h$ factors.  Given Eqs.~(\ref{eq:psj_Knum}) and (\ref{eq:psj_Ktherm}) and plausible assumptions about $b$ or $f_D$, it is possible to estimate the time scales for a wide variety of atomic and molecular collision processes under various kinds of conditions.

Now we will examine the important case of tunable resonant scattering when a closed channel is present.  Assume that open entrance channel $\alpha$, with $E_\alpha$ chosen as $E_\alpha=0$, is coupled through terms in the system Hamiltonian to a closed channel $\beta$ with $0 < E < E_\beta$.  Then a molecular bound state in channel $\beta$ becomes a quasibound state that acts as a scattering resonance in channel $\alpha$.  Using Fano's form of resonant scattering theory~\cite{Fano61}, let us assume a ''bare'' or uncoupled approximate bound state $|C\rangle=\psi_c(R)|c\rangle$ with energy $E_c$ in the closed channel $\beta=c$ and a "'bare'' or background scattering state $|E\rangle=\psi_{bg}(R,E)|bg\rangle$ at energy $E$ in the entrance channel $\alpha=bg$.  The scattering phase shift $ \eta(E)=\eta_{bg}(E)+\eta_{res}(E)$ of the coupled system picks up a resonant part due to the Hamiltonian coupling $W(R)$ between the "bare" channels.   Here $\eta_{bg}$ is the phase shift due to the uncoupled single background channel, as described in the last Section, and 
\begin{equation}
\label{eq:psj_etares}
\eta_{res}(E) = -\tan^{-1}\left ( \frac{\frac{1}{2}\Gamma(E)}{E-E_c-\delta E(E)}\right ) \,,
\end{equation}
has the standard Breit-Wigner resonance scattering form.  The two key features of the resonance are its width
\begin{equation}
\label{eq:psj_width}
  \Gamma(E) = 2 \pi |\langle C |  W(R) | E\rangle|^2 \,,
\end{equation}
and its shift
\begin{equation}
\label{eq:psj_shift}
 \delta E(E)= {\cal{P}}\int_{-\infty}^\infty \frac{|\langle C |  W(R) | E' \rangle|^2}{E-E'} dE' \,.
\end{equation}

The primary difference between an ''ordinary'' resonance and a threshold one as $E\to 0$ is that for the former we normally make the assumption that $\Gamma(E)$ and $\delta E(E)$ are evaluated at $E=E_c$ and are independent of $E$ across the resonance.  By contrast, the explicit energy dependence of $\Gamma(E)$ and $\delta E(E)$ are key features of threshold resonances~\cite{Bohn99,Julienne06,Marcelis04}.  In the special case of the $E\to 0$ limit for $s$-waves, 
\begin{eqnarray}
  \frac{1}{2} \Gamma(E) & \to & (ka_{bg})\Gamma_0  \label{eq:psj_Gamma0} \\
   E_c + \delta E(E) &  \to & E_0 \,,  \label{eq:psj_shift0}
\end{eqnarray}
where $\Gamma_0$ and $E_0$ are $E$-independent constants.  Note that $\Gamma(E)$ is positive definite, so that $\Gamma_0$ has the same sign
as $a_{bg}$. Assuming an entrance channel without inelastic loss, so that $ \eta_{bg}(E) \to -k a_{bg}$, and for the sake of generality, adding a decay rate $\gamma_c/\hbar$ for the decay
of the bound state $|C\rangle$ by irreversible loss processes, gives in the limit
of $E \to 0$,
\begin{equation}
 \tilde{a} = a-ib= a_{bg} - \frac{a_{bg}\Gamma_0}{E_0-i(\gamma_c/2)} \,.
      \label{eq:psj_ares}
\end{equation}
This formalism accounts for both  kinds of tunable resonances that are used for making cold molecules from cold atoms, namely, magnetically or optically tuned resonances.  We now give our attention to each of these in turn.

\section{Magnetically tunable resonances} \label{sec:psj_magres}

Cold alkali metal atoms have a variety of magnetically tunable resonances that have been exploited in a number of experiments to control the properties of ultracold quantum gases or to make cold molecules.  For the most part, experiments have succeeded with species that either do not have inelastic loss channels, or if they do, the loss rates are very small.  Thus, for practical purposes, we can set the resonance decay rate $\gamma_c=0$ in examining a wide class of magnetically tunable resonances.   While general coupled channel methods can be set up to solve the multichannel Schr{\"o}dinger equation~\cite{Kohler06}, we will use simpler models to explain the basic features of tunable Feshbach resonance states.

Many resonances occur for alkali metal species in their $^2$S electronic ground state because of their complex hyperfine and Zeeman substructure with energy splittings very large compared to $k_B T$.  Thus,  closed spin channels that have bound states near $E_\alpha$ of an entrance channel $\alpha$ can serve as tunable scattering resonances for threshold collisions in that channel.  The key to magnetic tuning of a resonance is that the resonance state $|C\rangle$ has a different magnetic moment $\mu_c$ than the moment $\mu_\mathrm{atoms}$ of the pair of separated atoms in the entrance channel.  The bare bound state energy can be tuned by varying the magnetic field $B$
\begin{equation}
\label{eq:psj_Ec}
 E_c(B)=\delta \mu(B-B_c) \,,
\end{equation}
where $\delta \mu = \mu_\mathrm{atoms}-\mu_c$ is the magnetic moment difference and $B_c$ is the field where $E_c(B_c)=0$ at threshold.  The scattering length is real with $b=0$ and takes on the following resonant form
\begin{equation}
\label{eq:psj_Ares}
 a(B) = a_{bg}-a_{bg}\frac{\Delta}{B-B_0}\,,
\end{equation}
where
\begin{equation}
 \Delta = \frac{\Gamma_0}{\delta \mu} \quad {\rm and}\quad 
 B_0=B_c+\delta B \,.
      \label{eq:psj_Delta}
\end{equation}
Note that the interaction between the entrance and closed channels shifts the point of singularity of $a(B)$ from $B_c$ to $B_0$.  Such magnetically tunable Feshbach resonances are characterized by four parameters, namely, the background scattering length $a_{bg}$. the magnetic moment difference $\delta \mu$, the resonance width $\Delta$, and position $B_0$.

\begin{figure}[tbp]
\centering
\includegraphics[width=\columnwidth]{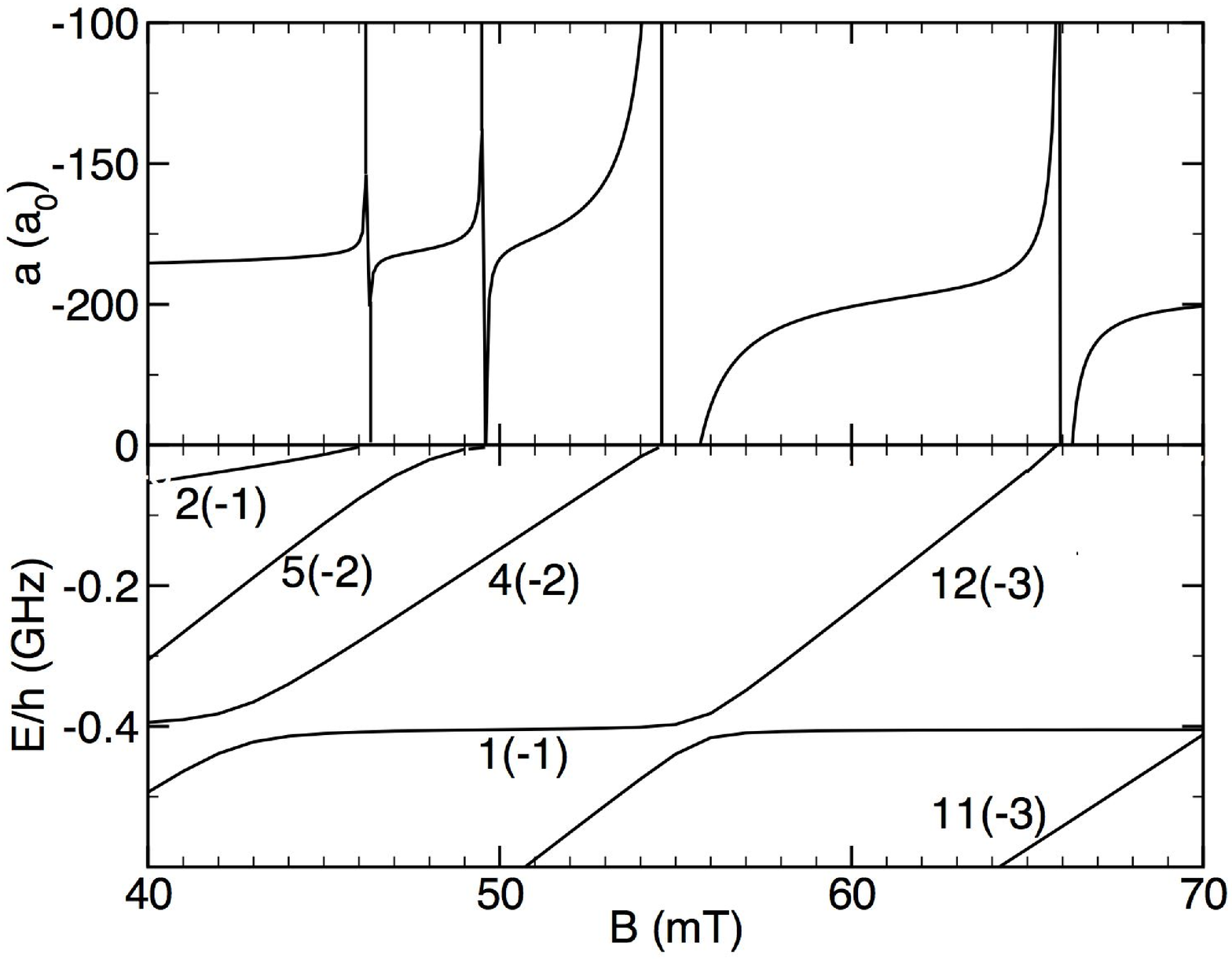}
\caption{Molecular bound state energies (lower panel) and scattering length (upper panel) versus magnetic field $B$ in mT (1 mT $=$ 10 Gauss) for the lowest energy $\alpha=1$ $s$-wave spin channel of the $^{40}$K$^{87}$Rb fermionic molecule.  The bound state energies are shown relative to the channel energy $E_1$ of the two separated atoms taken to be zero.  This $\alpha=1$ spin channel has respective $^{40}$K and $^{87}$Rb spin projection quantum numbers of $-9/2$ and $+1$, giving a total projection of $-7/2$.  In this species there are 11 additional closed $s$-wave channels with $E_\alpha>E_1$ and with the same projection of $-7/2$.  The bound state quantum numbers are $\alpha(i)$, where $i$ is the vibrational quantum number relative to the dissociation limit of closed channel $\alpha = 2,\ldots,12$. Four bound states cross threshold in this range of $B$, giving rise to singularities in the scattering length.  }
\label{fig:psj6}
\end{figure}

Figure~\ref{fig:psj6} shows an example of the scattering length and bound state energies for the $^{40}$K$^{87}$Rb molecule near the lowest energy spin channel of the separated atoms.  The spin quantum numbers and hyperfine splitting in their respective electronic ground states are 1, 2, and 6.835 GHz for $^{87}$Rb and $9/2$, $7/2$ and $-1.286$ GHz (inverted) for $^{40}$K.  There are 11 other closed  spin channels in this system with $E_\alpha >E_1$  that have the same total projection quantum number as the lowest energy $\alpha=1$ spin channel.  Because of their different magnetic moments the energy of a bound state of one of these closed channels can be tuned relative to the energy of the two separated atoms in the $\alpha=1$ $s$-wave channel, as shown in the Figure.  Due to coupling terms in the Hamiltonian among the various channels, bound states that cross threshold couple to the entrance channel and give rise to resonance structure in its $a(B)$.  The resonance with $B_0$ near 54.6 mT (546 G) has been used to associate a cold $^{40}$K atom and a cold $^{87}$Rb atom to make a $^{40}$K$^{87}$Rb molecules in a near-threshold state with a small binding energy on the order of 1 MHz or less~\cite{Ospelkaus06}. 

It is extremely useful to introduce the properties of the long range van der waals potential and take advantage of the separation of short and long range physics discussed in the previous Section.  Assuming that the interaction $W(R)$ is confined to distances $R \ll \bar{a}$, the matrix element in Eq.~(\ref{eq:psj_width}) defining $\Gamma(E)$ can be factored as
\begin{equation}
 \Gamma(E) = C_{bg,\ell}(E)^{-2} \bar{\Gamma} \,,
 \label{eq:psj_factored}
\end{equation}
where $\bar{\Gamma}$ is a measure of resonance strength that depends only on the energy-independent short-range physics near $E=0$, and is completely independent of the asymptotic boundary conditions.   It thus can be used in characterizing the properties of both scattering and bound states when $E \ne 0$.

The extrapolation of resonance properties away from $E=0$ depends on two additional parameters associated with the long range potential, $\mu$ and $C_6$, which determine $\bar{a}$ and $\bar{E}$.  Let us define a dimensionless resonance strength parameter
\begin{equation}
\label{eq:psj_sres}
 s_{res}=\frac{a_{bg}\delta \mu \Delta}{\bar{a}\bar{E}} = r_{bg}\frac{\Gamma_0}{\bar{E}} \,
\end{equation}
where $r_{bg}=a_{bg}/\bar{a}$.  Using the threshold van der Waals form of $C_{bg,0}(E)^{-1}$ given in the previous section, we can write
\begin{equation}
\label{eq:psj_Gammabar} 
\frac{\bar{\Gamma}}{2} = (s_{res}\bar{E})\frac{1}{1 + (1-r_{bg})^2} \,.
\end{equation}
The above-threshold scattering properties are found from the scattering phase shift $\eta(E)=\eta_{bg}(E)+\eta_{res}(E)$, where $\eta_{res}(E)$ is found from Eq.~(\ref{eq:psj_etares}) once $E_c$, $\Gamma(E)$ and $\delta E(E)$ are known.  The first two are given by Eqs.~(\ref{eq:psj_Ec}) and (\ref{eq:psj_factored}), and 
\begin{equation}
\label{eq:psj_vdwshift}
 \delta E(E) = \frac{\bar{\Gamma}}{2} \tan{\lambda_{bg}(E)},\,
\end{equation}
where $\tan{\lambda_{bg}(E)}$ is a function determined by the van der Waals potential, given $a_{bg}$.  It has the limiting form $\tan{\lambda_{bg}(E)} =1-r_{bg}$ as $E\to 0$, and $\tan{\lambda_{bg}(E)=0}$ for $E \gg \bar{E}$~\cite{Julienne89,Mies00}.  Thus the position of the scattering length singularity is shifted by 
\begin{equation}
\label{eq:psj_Bshift}
 \delta B = B_0 - B_c = \Delta \frac{r_{bg}(1-r_{bg})}{1+(1-r_{bg})^2}
\end{equation}
from the crossing point $B_c$ of the ``bare'' bound state.  Scattering phase shifts calculated from the van der Waals potential with the``quantum defect" forms in Eqs.~(\ref{eq:psj_factored}) and (\ref{eq:psj_vdwshift}) are generally in excellent agreement with complete coupled channels methods for energy ranges on the order of $\bar{E}$ and even larger~\cite{Julienne06}.

The properties of bound molecular states near threshold can also be calculated from the general coupled-channels quantum defect method using the properties of the long range potential.  When the energy $E_b(B)=-\hbar^2k_b(B)^2/(2\mu)$ of the threshold $s$-wave bound state is small, that is, $|E_b(B)| \ll \bar{E}$ or $k_b(B) \bar{a} \ll 1$, then the equation for $E_b(B)$ from the quantum defect method is
\begin{equation}
\label{eq:psj_qdtlastE}
\left (E_c(B) -E_b(B)\right ) \left ( \frac{1}{r_{bg}-1} - k_b(B) \bar{a} \right ) = \frac{\bar{\Gamma}}{2} \,.
\end{equation}
If $\bar{\Gamma} =0$, we recover the uncoupled, or ``bare,'' bound states of the system, whereas when $\bar{\Gamma}>0$, this equation gives the coupled, or ``dressed,'' bound states.  The threshold bound state ``disappears'' into the continuum at $B=B_0$, where $a(B)$ has a singularity. The shift in Eq.~\ref{eq:psj_Bshift} follows immediately upon solving for $E_c(B_0)$ where $E_b(B_0)=0$. 

Threshold bound state properties are strongly affected by the magnitudes of $s_{res}$ and $r_{b g}$.  When the coupled bound state wavefunction is expanded as a mixture of closed and background channel components, $|c\rangle$ and $|bg\rangle$ respectively, an important property is the norm $Z(B)$ of the closed channel component; the norm of the entrance channel component is $1-Z(B)$.  The value of $Z$ can be calculated from a knowledge of $E_b(B)$, since $Z=|\delta \mu^{-1}\partial E_b/\partial B|$~\cite{Kohler06}.  

There are two basic classes of resonances.  One, for which $s_{res}\gg1$, are called entrance channel dominated resonances.  These have $Z(B)\ll 1$ as $B-B_0$ varies over a range that is a significant fraction  of $|\Delta|$.  In addition, the bound state energy is given by Eq.~(\ref{eq:psj_vdwE}) over a large part of this range.  On the other hand, closed channel dominated resonances are those with $s_{res} \ll 1$.  They have $Z(B)$ large, on the order of unity, as $|B-B_0|$ varies over a large fraction of $|\Delta|$, and only have a "universal" bound state Eq.~(\ref{eq:psj_universalE})  over a quite small range $\ll |\Delta|$ near $B_0$.  Entrance channel dominated resonances have $\Gamma(E,B) > E$ when $0 < E< \bar{E}$, so that no sharp resonance feature persists above threshold, where $a(B) <0$ and the last bound state has disappeared.  By contrast, closed channel dominated resonances with $|r_{bg}|$ not too large  will have $\Gamma(E,B) < E$ when $0 <E<\bar{E}$, so that a sharp resonant feature emerges just above threshold, continuing as a quasibound state with $E>0$ into the region where $a(B)<0$.

\begin{figure}[tbp]
\centering
\includegraphics[width=\columnwidth]{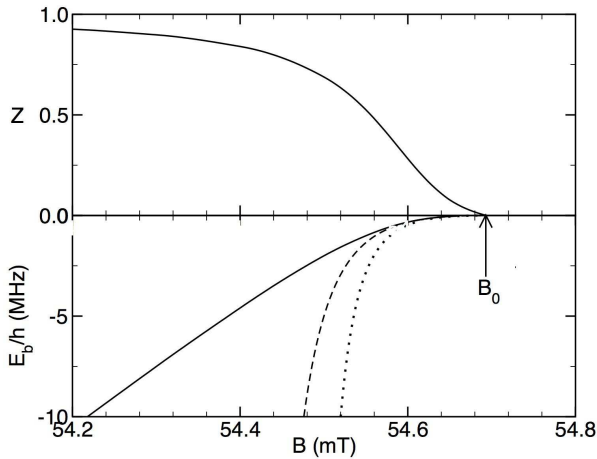}
\caption{The lower panel shows an expanded view of $E_b(B)$ near $B_0$ for the $^{40}$K$^{87}$Rb resonance with $B_0=54.693$ mT (546.93 G) in Fig.~\ref{fig:psj6}.  The solid line comes from a coupled channels calculation that includes all 12 channels with the same $-7/2$ projection quantum number. The dashed and dotted lines respectively show the universal energy of Eq.~(\ref{eq:psj_universalE}) and the van der Waals corrected energy of Eq.~(\ref{eq:psj_vdwE}).  The upper panel shows the closed channel norm $Z(B)$.  The width $\Delta=0.310$ mT (3.10 G), $a_{bg}=-191$ a$_0$, and $\delta \mu/h=33.6$ MHz/mT (3.36 MHz/G).  With $\bar{a}=68.8$ a$_0$ and $\bar{E}/h=13.9$ MHz, this is a marginal entrance channel dominated resonance with $s_{res}=2.08$.}
\label{fig:psj7}
\end{figure}

\begin{figure}[tbp]
\centering
\includegraphics[width=\columnwidth]{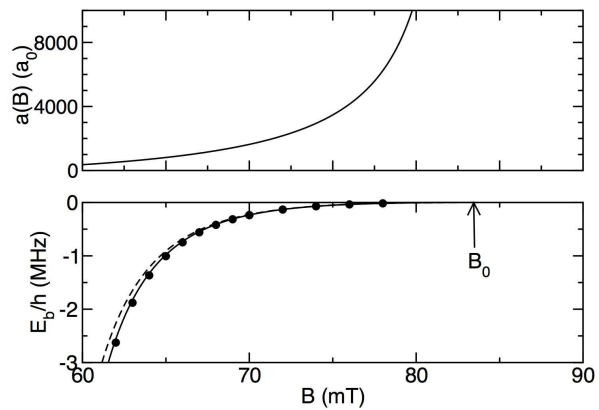}
\caption{Molecular bound state energy (lower panel) and scattering length (upper panel) versus magnetic field $B$ for the lowest energy $\alpha=1$ $s$-wave spin channel of the $^{6}$Li$_2$ molecule.  This channel has one $^6$Li atom in the lowest  $+1/2$ projection state and the other in the lowest $-1/2$ projection state for a total projection of $0$.  There are 4 additional closed channels with projection $0$.  In this range of $B$ there is only one bound state that crosses threshold at $B_0=83.4$ mT (834 G).  The lower panel shows $E_b(B)$ from a coupled channels calculation (solid circles), the universal limit of Eq.~(\ref{eq:psj_universalE}) (dashed line) and the corrected limit of Eq.~(\ref{eq:psj_vdwE}) (solid line).   The width $\Delta=30.0$ mT (300 G),  $a_{bg}=-1405$ a$_0$, and $\delta \mu/h=28$ MHz/mT (2.8 MHz/G).  This is a strongly entrance channel dominated resonance with $s_{res}=59$, and $Z<0.06$ over the range of $B$ shown.}
\label{fig:psj8}
\end{figure}

Figure~\ref{fig:psj7} shows an expanded view of the $4(-2)$ resonance of $^{40}$K$^{87}$Rb near 54.6 mT.  The figure shows the character of the bound state as it merges into threshold at $B_0$.  It tends to be a universal ``halo'' bound state over a range of $|B-B_0|$ that is less than about $1/3$ of $\Delta$.  As $|B-B_0|$ increases, the bound state increasingly takes on the character of the closed channel $4(-2)$ level as $Z$ increases towards unity.  Figure~\ref{fig:psj8} shows an example of the very broad $^6$Li resonance in the lowest energy $\alpha=1$ $s$-wave channel, which requires two $^6$Li fermions in different spin states.  This is a strongly entrance channel dominated resonance, where $Z\ll 1$ over a range of $|B-B_0|$ nearly as large as $\Delta$.  The last bound state is a universal halo molecule over a range larger than 100 G.  The corrected Eq.~(\ref{eq:psj_vdwE}) is a good approximation over an even larger range.  The scattering length graph shows that the size $\approx a(B)/2$ of the halo state is very large compared to $\bar{a}=30$ a$_0$ (see Table~\ref{tab:psj_vdw}) over this range.

Magnetically tunable scattering resonances have proven very useful in associating two cold atoms to make a molecule in the weakly bound states near threshold.  This work is reviewed in detail in Ref.~\cite{Kohler06}.  The magneto-association process works by first preparing a gas with a mixture of both atomic species at $B>B_0$ (assuming $\delta \mu >0$), where there is no threshold bound state.  By ramping the $B$ field down in time so that $B<B_0$, colliding pairs of atoms with $E>0$ can be converted to diatomic molecules in a bound state with energy $E<0$.   The conversion efficiency will depend on both the ramp rate and the phase space density of the initial gas.    If the initial atom pair is held in a single cell of an optical lattice instead of a gas, the conversion efficiency can approach 100 per cent.  A simple Landau-Zener picture has been found to be quite accurate for such lattice cells, where the conversion probability of the atom pair in the trap ground state $i=0$ is  $1-e^{-A}$, where 
\begin{equation}
\label{eq:psj_LZA}
 A = \frac{2\pi}{\hbar} \frac{W_{ci}^2}{\dot{E_c}} \,.
\end{equation}
Here $i \ge 0$ represents the above-threshold levels of the atom pair confined by the trap, continuing the below threshold series of dimer levels with $i\le -1$.  
For a three dimensional harmonic trap with frequency $\omega_x=\omega_y=\omega_z=\omega$, the matrix element $W_{ci} = \langle C|W(R)|i\rangle$ is well-approximated as $W_{ci}=\sqrt{\Gamma(E_i)/2\pi}\sqrt{\partial E_i/\partial i}$, where $\partial E_i/\partial i=2\hbar \omega$ and $\Gamma(E_i)=2k_i a_{bg}\delta \mu \Delta$ for the $i=0$ trap ground state of relative motion with $k_i=\sqrt{3\mu\omega/\hbar}$ (see Eqs.~\ref{eq:psj_width}, ~\ref{eq:psj_Gamma0} and~\ref{eq:psj_Delta}).  The trick used here in getting a matrix element $W_{ci}$ between two bound states from the matrix element  $ \langle C|W(R)|E\rangle$ involving an energy-normalized scattering state is to introduce the density of states as in Eq.~(\ref{eq:psj_psi-}).  In a similar manner, the matrix element can be obtained between the bare closed channel state and the bound states $i<0$ of the entrance channel.  Such matrix elements characterize avoided crossings like the one  in Fig.~\ref{fig:psj6} for $E/h$ near $-0.4$ MHz and $B$ near 43 mT.  Finally, it should be noted that a Landau-Zener model can also be used for molecular dissociation by a fast magnetic field ramp.  An alternative phenomenological model has been developed to describe molecular association in cold gases, which are more complex than two atoms in a lattice cell~\cite{Hodby05}.

\section{Photoassociation} \label{sec:psj_optres}

\begin{figure}[tbp]
\centering
\includegraphics[width=\columnwidth]{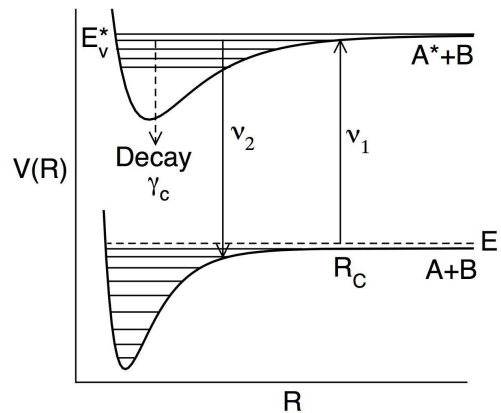}
\caption{Schematic representation of one- and two-color photoassociation (PA).  The two colliding ground state atoms at energy $E$ can absorb a laser photon of frequency $\nu_1$ and be excited to an excited molecular bound state at energy $E_v^*$.  The bound state decays via spontaneous emission at rate $\gamma_c/\hbar$.  If a second laser is present with frequency $\nu_2$, the excited level can also be coupled to a ground state vibrational level $v$ at energy $E_v$, if $h(\nu_2-\nu_1)=E-E_v$.  The PA process depends on the ground state wavefunction at the Condon point $R_C$ of the transition, where $h\nu_1$ equals the difference between the excited and ground state potentials.}
\label{fig:psj9}
\end{figure}

Cold atoms can also be coupled to molecular bound states through photoassociation (PA), as discussed in Chapters {\color{red} XX, YY, and ZZ}.    Fig.~\ref{fig:psj9} gives a schematic description of PA, a process by which the colliding atoms can be coupled to such bound state resonances through one- or two-photons.   Reference~\cite{Jones06} reviews theoretical and experimental work on PA spectroscopy and molecule formation.  Molecules made using the magnetically tunable resonances described in the last Section are necessarily very weakly bound, with binding energies limited by the small range of magnetic tuning.   Photoassociation has the advantage that laser frequencies are widely tunable, so that a range of many bound states becomes accessible to optical methods, even the lowest $v=0$ vibrational level of the ground state.  In addition, the light can be turned off and on or varied in intensity for time-dependent manipulations.

Photoassociation naturally lends itself to the resonant scattering treatment of a decaying resonance in Eq.~(\ref{eq:psj_ares}), which applies to the one-color case with position $E_c = E_v^* - h \nu_1$, strength $a_{bg}\Gamma_0(I) = \Gamma(E,I)/(2k)$, and shift $\delta E(I)$. The latter two are linear in laser intensity $I$ when $I$ is low enough.  PA is usually detected by the inelastic collisional loss of cold atoms it causes, due to the spontaneous decay of the excited state to make hot atoms or deeply bound molecules.  In the limit $E \to 0$ the complex scattering length is
\begin{eqnarray}
a(\nu_1,I)  & = & a_{bg}-L_{opt}\frac{\gamma_c E_0}{E_0^2+(\gamma_c/2)^2} \\
b(\nu_1,I) & = & \frac{1}{2}L_{opt}\frac{\gamma_c^2}{E_0^2+(\gamma_c/2)^2} \,,
\end{eqnarray}
where $E_0=E_v^* -h\nu_1+\delta E(I)$ is the detuning from resonance, including the intensity-dependent shift, and the optical length is defined by $L_{opt}=a_{bg}\Gamma_0(I)/\gamma_c$.  

Photoassociation spectra, line shapes, and shifts have been widely studied for a variety of like and mixed alkali-metal species.  At the higher temperatures often encountered in magneto-optical traps, contributions to PA spectra from higher partial waves, e.g., $p$- or $d$-waves, have been observed in a number of cases.  The theory can be readily extended to higher partial waves.  By introducing an energy-dependent complex scattering length the theory for $s$-waves can be extended to finite $E$ away from threshold and to account for effects due to reduced dimensional confinement in optical lattices~\cite{Naidon06}.  

The optical length formulation of resonance strength is very useful for a decaying resonance.  It also applies to decaying magnetically tunable resonances, if $\Gamma_0$ from Eq.~(\ref{eq:psj_Delta}) is used to define a resonance length $a_{bg}\delta \mu \Delta/\gamma_c$ equivalent to $L_{opt}$~\cite{Hutson07}.  The scattering length has its maximum variation of $a_{bg} \pm L_{opt}$ when the laser is tuned to $E_0=\pm\gamma_c/2$, and losses are maximum at $E_0=0$ where $b=L_{opt}$.  When detuning is small, on the order of $\gamma_c$, significant changes to the scattering length on the order of $\bar{a}$ are thus normally accompanied by large loss rates (see Eq.~\ref{eq:psj_Knum}).     Losses can be avoided by going to large detuning, since when $( \gamma_c/E_0) \ll 1 $, $b = (L_{opt}/2)(\gamma_c/E_0)^2$, whereas the change in $a$ only varies as $a-a_{bg}=-L_{opt}(\gamma_c/E_0)$. To make the change $a-a_{bg}$ large enough while requiring $(\gamma_c/E_0)\ll1$ means that $L_{opt}$ has to be very large compared to $\bar{a}$.

The magnitude of $L_{opt}$ depends on the matrix element $\langle C|\hbar \Omega_1(R) | E \rangle$ where $\hbar \Omega_1(R)$ represents the optical coupling between the ground and excited state.  Using Eqs.~(\ref{eq:psj_width}) and (\ref{eq:psj_Gamma0}) and the above definition of $L_{opt}$, and factoring out the relatively constant $\hbar \Omega_1$ value, 
\begin{equation}
\label{eq:psj_Lopt}
 L_{opt} = \pi \frac{|\hbar \Omega_1|^2}{\gamma_c} \frac{F(E)}{k} \,.
\end{equation}
The Franck-Condon overlap factor is
\begin{eqnarray}
F(E) & = &\left | \int_0^\infty \psi_v^*(R) \psi_0(R,E) dR \right |^2 \label{eq:psj_FCF}  \\
 & \approx & \frac{\partial E_v^*}{\partial v} \frac {1}{D_C} |\psi_0(R_C,E)|^2  \,, \label{eq:psj_refapprox}
\end{eqnarray}
where $D_C$ is the derivative of the difference between the excited and ground state potentials evaluated at the Condon point $R_C$, and  ${\partial E_v^*}/{\partial v}$ is the excited state vibrational spacing.
Equation~(\ref{eq:psj_refapprox}) is known as the reflection approximation, generally an excellent approximation where $F(E)$ is proportional to the square of the ground state wavefunction at $R_C$, the Condon point where the molecular potential difference matches $h\nu_1$ (see Fig.~\ref{fig:psj9}).  Thus, $F(E)$ can be evaluated using expressions like Eqs.~(\ref{eq:psj_psi}) or (\ref{eq:psj_psiE+}) for $R_C\gg \bar{a}$ or $R_C\ll \bar{a}$ respectively.   The reflection approximation is quite good over a wide range of $E$ and for higher partial waves than the $s$-wave.  By selecting a range of excited levels $v$ by changing laser frequency $\nu_1$, thus changing $R_C$, the shape and nodal structure of the ground state wavefunction can be mapped out over a range of $R$.

The optical length has several important properties evident from Eq.~(\ref{eq:psj_Lopt}).  First, since both $\Omega_1$ and $\gamma_c$ are proportional to the same squared transition dipole moment, $L_{opt}$ does not depend on whether the transition is strong or weak, but can be large for both kinds of transitions.  Second, $L_{opt} \propto |\Omega_1|^2$ so it can be increased by increasing laser intensity.  Third, since $F(E) \propto k$ as $E \to 0$ for entrance channel $s$-waves, $L_{opt} \propto F(E)/k$ is independent of $E$ or $k$ at low energy.  However, it does depend strongly on the molecular structure through the Franck-Condon factor.  In practice, using strong transitions with large decay rates such as those in alkali-metal species leads to the requirement to use excited molecular levels far from threshold with large binding energies.  This is necessary to achieve large detuning from atomic and molecular resonance.  This requirement means such levels have small $F(E)$ factors, due to the very large value of $D_C$ in Eq.~(\ref{eq:psj_refapprox}).  On the other hand, weak transitions with small decay rates, such as those associated with the $^1$S$_0 \to ^3$P$_1$ intercombination line transition for alkaline earth species such as Sr, can lead to quite large values of $L_{opt}$.  This is because large detuning in $\gamma_c$ units can be achieved for levels that are still quite close to the excited state threshold.  Such levels typically have large Franck-Condon factors.  In fact, PA transitions near the weak intercombination line of Sr have been observed to have  $L_{opt}$ several orders of magnitude larger than was observed for strongly allowed molecular transitions involving Rb~\cite{Zelevinsky06}.  Thus, there are good prospects for some degree of optical resonant control of collisions in ultracold gases of species like Ca, Sr, or Yb.
 
Two-color PA is also possible when a second laser with frequency $\nu_2$ is added, as shown in Fig.~\ref{fig:psj9}.  When the the frequency difference is chosen so that $h(\nu_2-\nu_1)=E-E_{i\ell}$, the ground $i\ell$ molecular level is in resonance with collisions at energy $E$.  By keeping $\nu_1$ fixed and tuning $\nu_2$, two-color PA spectroscopy can be used to probe the level energies.  This is how the data on binding energies of the Yb$_2$ molecule was obtained~\cite{Kitagawa07} so as to be able to construct Fig.~\ref{fig:psj5}.  In this case, 12 different levels from different isotopic species were measured, among which were levels with  $i=-1$ and $-2$ and $\ell=0$ and $2$.  Two color spectroscopy has also been carried out for several alkali-metal homonuclear species.  

Two-color processes are also an excellent way to assemble two cold atoms into a translationally cold molecule.  Early work along these lines was done using the spontaneous decay of the excited level to populate a wide range of levels in the ground state.  The disadvantage of spontaneous decay is that it is not selective.  However, by using a laser with a precise frequency, a specific level can be chosen as the target level.  One early experiment did this to associate two $^{87}$Rb atoms in a Bose-Einstein condensate to make a molecular level at a specific energy of $h(-636)$ MHz~\cite{Wynar00}.  

It is highly desirable to be able to make translationally cold molecules in their vibrational ground state $v=0$.  This is especially true of polar molecules, which have large dipole moments in $v=0$.  On the other hand, threshold levels have negligible dipole moments, since there is no charge transfer because of the large average atomic separation $\approx \bar{a}$.  A promising technique is to use magnetoassociation using a tunable Feshbach resonance to associate the atoms into a threshold molecular level, then use a 2-color Raman process to move the population in that state to a much more deeply bound level.  Although molecules in a gas are subject to fast destructive collisions with cold atoms or other molecules in the gas (see Eq.~\ref{eq:psj_Knum}), the molecules can be protected against such collisions by forming them in individual optical lattice trapping cells.  Then the 2-color Raman process could be used to produce much more deeply bound molecules that are stable against destructive collisions.  This has been done successfully with $^{87}$Rb$_2$~\cite{Winkler07} molecules.  In the future, such methods are likely to produce $v=0$ polar molecules, with which a range of interesting physics can be explored~\cite{Lewenstein06,Buchler07}.


\bibliographystyle{apsrev}

\bibliography{bib_psj}

\end{document}